\documentclass[twocolumn,showpacs,preprintnumbers,amsmath,amssymb,aps,prb]{revtex4}

\usepackage{graphicx}
\usepackage{dcolumn}% Align table columns on decimal point
\usepackage{bm}% bold math

\begin{document}

\title{
Statistics of  
orbital entanglement production in quantum-chaotic dots}
\author{Victor A. Gopar}
\affiliation{Departamento de F\'isica Te\'orica and Instituto de Biocomputaci\'on y F\'isica de Sistemas Complejos, Universidad de Zaragoza, Pedro Cerbuna 12, E-50009 Zaragoza, Spain}

\author{Diego Frustaglia}
\affiliation{Departamento de F\'isica Aplicada II, Universidad de Sevilla, E-41012 Sevilla, Spain}

%\date{\today}

\begin{abstract}

The production of orbitally entangled electrons in quantum-chaotic dots is investigated from a statistical point 
of view. The degree of entanglement is quantified through the concurrence and the entanglement of formation. 
We calculate the complete statistical distributions of the entanglement measures by using random matrix theory. Simple analytical expressions are provided for the concurrence distributions. We identify clear signatures of time-reversal invariance in the production of entanglement at the level of the entanglement-measure distributions, such as the ability of producing maximally entangled (Bell) states, which passed unnoticed in previous works 
where only the first two moments of the distributions were studied.    

\end{abstract}


\maketitle

In the early times of quantum mechanics, the prediction of entangled quantum states subject to  
nonclassical correlations was identified as a possible sign of incompleteness of the theory.\cite{EPR35} 
Despite such original objections, and after many years of debate during which decisive experimental 
evidence was accumulated, the reality of quantum entanglement is nowadays widely accepted. Besides 
its fundamental interest, modern research on entanglement has been strongly motivated by its possible application  
as a resource for quantum information processing. \cite{NC-book} As a consequence, an intensive research 
activity on the production, manipulation, and detection of quantum entanglement in a variety of physical systems is reflected in an extended literature. 

Several efforts have been devoted to the study of the entanglement in electronic systems\cite{B06,B07} with 
a view on solid-state applications that could lead to a major technological breakthrough in the field of nanoelectronics. 
As far as the production 
of entangled electrons is concerned, several proposals based on different interacting\cite{inter} and noninteracting\cite{noninter,
FMF06,BKMY04} electron mechanisms already exist. 
Recently, Beenakker et. al.\cite{BKMY04} proposed a ballistic quantum dot as an orbital entangler for 
pairs of noninteracting electrons\cite{samuelsson}; by assuming that the entangler is a chaotic quantum dot, they calculated the average and variance of the concurrence (a standard measure of two-qubit entanglement) and found that these two moments 
are practically unaffected by the breaking of time-reversal invariance (TRI). This fact is in  
contrast to other transport properties of ballistic quantum dots such as the conductance, where TRI yields 
significant weak-localization corrections. More recently, it was shown\cite{FMF06} that signatures of 
TRI corrections can arise in the average concurrence 
once spin-orbit interaction is considered. The concurrence fluctuations, however, can be very large (of the order of 
the average concurrence).\cite{FMF06} This indicates that the first two moments of the concurrence are 
insufficient for an accurate statistical description of the entanglement production in a quantum dot.

Here we calculate the complete distribution of the degree of orbital entanglement in terms of the 
concurrence $\mathcal C$ and the entanglement of formation $\mathcal{E}$. We derive simple analytical  
expressions for the distribution of $\mathcal C$ for the cases when TRI is preserved and broken, while 
the distributions  of $\mathcal{E}$ are calculated via a relation between $\mathcal C$ and $\mathcal{E}$.
We find that the degree of produced entanglement is distributed quite differently depending on whether TRI is present or not. 
In particular, we show that maximally entangled states are produced only when TRI is preserved. 
Clear signatures of TRI correlations are, however, absent in the first two moments of the distributions, which is 
in agreement with previous results.\cite{BKMY04} We verify our theoretical results for the concurrence distributions by numerical simulations of chaotic quantum dots.

\begin{figure}
\includegraphics[width=0.75\columnwidth]{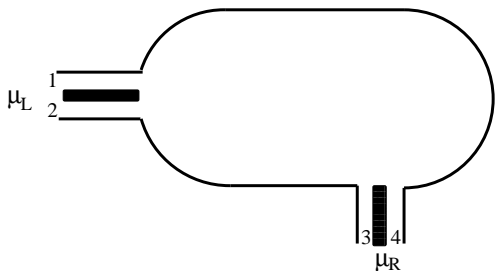}
\caption{The quantum dot orbital entangler. The left and right leads are attached to chemical potentials $\mu_L$ and $\mu_R$, respectively. As explained in the text, an electron leaving the dot to the left (right) side can escape through the leads 1 or 2 
(3 or 4), defining a two-level quantum system.}
\label{fig-1}
\end{figure}

The setup of the orbital entangler as proposed in Ref.~\onlinecite{BKMY04} is sketched in Fig.~\ref{fig-1}. 
It consists of a quantum dot with two attached single-channel leads at the left and  right. Each lead is 
connected to an electron reservoir. The application of a small bias voltage between reservoirs give rise to a 
coherent current traversing the dot from left to right. Exchange correlations due to scattering within the 
dot lead to entanglement between transmitted (to the right) and reflected (to the left) electrons, as we show in the following. 

We start by considering a separable two-particle state incoming from the left reservoir 
in Fig.~\ref{fig-1}:
\begin{equation}
|\Psi_{\rm in}\rangle=a_1^\dag a_2^\dag |0\rangle ,
\label{Psi-in}
\end{equation}
where $a_i^\dag$ creates an incoming electron in the lead $i=1$, and $2$ and $|0\rangle$ denotes the Fermi sea at 
zero temperature. We disregard spin degeneracy for simplicity (equivalently, one can consider incoming 
electrons with their spins polarized along the same direction). The outgoing state is a coherent superposition 
of orbital channels determined by the single-particle scattering matrix $S$, such that 
\begin{equation}
|\Psi_{\rm out}\rangle=\sum_{mn} S_{m1} S_{n2} b_m^\dag b_n^\dag |0\rangle.
\label{Psi-out}
\end{equation}
Here, $b_j^\dag$ creates an outgoing electron in lead $j$, while the matrix element $S_{ji}$ 
describes the  scattering of an incident state from the lead $i (=1$ and $2)$ to the lead $j (=1,\dots,4)$.  Notice that the terms with $m=n$ in Eq.~(\ref{Psi-out}) vanish for the sake of 
fermionic statistics. 
The state $|\Psi_{\rm out}\rangle$ can be split into three components representing sectors of the Fock space with 
different local particle number at the left ($n_{\rm L}$) and right ($n_{\rm R}$) of the dot such that the total particle 
number remains constant 
($n_{\rm L}+n_{\rm R}= 2$):
\begin{equation}
|\Psi_{\rm out}\rangle=\sum_{n_{\rm L},n_{\rm R}} |n_{\rm L},n_{\rm R}\rangle=|2,0\rangle+|0,2\rangle+|1,1\rangle,
\label{Psi-out-n}
\end{equation}  
with
\begin{eqnarray}
\label{Psi-20}
|2,0\rangle&\equiv&(S_{31} S_{42}-S_{41} S_{32}) b_3^\dag b_4^\dag |0\rangle,\\
\label{Psi-02}
|0,2\rangle&\equiv&(S_{11} S_{22}-S_{21} S_{12}) b_1^\dag b_2^\dag |0\rangle,\\
\label{Psi-11}
|1,1\rangle&\equiv&\sum_{pq} (S_{p1} S_{q2}-S_{q1} S_{p2}) b_p^\dag b_q^\dag |0\rangle,
\end{eqnarray} 
where $p=1,2$ and $q=3,4$ in Eq.~(\ref{Psi-11}).
We intend to characterize the degree of orbital entanglement between the left and right outgoing channels. 
In practice, this could be addressed by studying current correlations in Bell-like measurements involving local 
transformations at the left and right sides of the dot.\cite{BKMY04,B06,faoro,witness} Such transformations acting on the state 
$|\Psi_{\rm out}\rangle$ conserve the local particle number, forbidding a mixing between the different terms of 
Eq.~(\ref{Psi-out-n}). Under these conditions, the total accessible entanglement\cite{WV03} is determined by 
a sum of distinct contributions from each individual sector of the Fock space. In our case, the sector 
contributing to the orbital entanglement is the one with equal occupancy at both sides of the dot (i.e., $|1,1\rangle$)   
since $|2,0\rangle$ and $|0,2\rangle$ are clearly separable in terms of the bipartition left-right.\cite{BKMY04,FMF06,B06} 
An electron leaving the quantum dot to the left side can choose between the leads $1$ and $2$ for escaping (see Fig.~\ref{fig-1}). 
This defines a two-level 
quantum system or qubit. The same happens with an electron escaping to the right side through leads $3$ and $4$. 
This means that the component $|1,1\rangle$ in Eq. (\ref{Psi-11}) describes (up to a normalization factor) 
a two-qubit entangled state. 

A widely used measure for quantifying two-qubit entanglement is the concurrence $\mathcal C$. It is defined as\cite{W98}
\begin{equation}
\mathcal C(\rho)\equiv {\rm max}\{0,\lambda_1-\lambda_2-\lambda_3-\lambda_4\},
\label{C}
\end{equation}
where the $\lambda_i$'s are the eigenvalues (in decreasing order) of the matrix $\rho \tilde{\rho}$, with $\rho$ being a 
$4\times 4$ two-qubit density matrix ($\rho=|1,1\rangle \langle1,1|/\langle1,1|1,1\rangle$ in our case) and  
$\tilde{\rho}=(\sigma_y \otimes \sigma_y)\rho^*(\sigma_y \otimes \sigma_y)$, where $\sigma_y$ the second Pauli matrix. The concurrence varies from 0 to 1. 
The case ${\mathcal C}=0$ corresponds to separable nonentangled states, while maximally entangled (Bell) states 
own $\mathcal C=1$. Those states with a $0 < \mathcal C < 1$ are non-separable partly entangled states. For our quantum dot entangler in Fig. 1, a finite value of $\mathcal C$ would guarantee that the left and right outgoing channels are orbitally entangled. 

The concurrence has the advantage of being directly related to the entanglement of formation $\mathcal E$, which is one of 
the most accepted measures of entanglement. Physically, $\mathcal E$ quantifies the cost, in terms of Bell states, to 
prepare 
a given (pure) state. The entanglement of formation and the concurrence are related through \cite{BKMY04,B06,W98}
\begin{equation}
\label{formation}
 \mathcal E (\mathcal C) = h\left( \frac{1+\sqrt{1-\mathcal C^2}}{2} \right),
\end{equation}
where
\begin{equation}
\label{h}
 h(x)=-x \log_2 (x) - (1-x)\log_2(1-x) .
\end{equation}

On the one hand,  $\mathcal C$ and $\mathcal E$
can be written in terms of the scattering matrix $S$ through the transmission eigenvalues $\tau_1$ and $\tau_2$ as\cite{BKMY04,B06}
\begin{equation}
\label{concu_t1t2}
\mathcal C=\frac{2\sqrt{\tau_1(1-\tau_1)\tau_2(1-\tau_2)}}{\tau_1+\tau_2-2\tau_1\tau_2} .
\end{equation}
We note that the entanglement is maximum ($\mathcal C=1$) when $\tau_1 =\tau_2$, 
and minimum ($\mathcal C=0$) when $\tau_1=0$ and $\tau_2=1$ or $\tau_1=1$ and $\tau_2=0$. 
The scattering matrix for the quantum dot of Fig.~\ref{fig-1} reads
\begin{equation}
S=\left[
\begin{array}{cc}
r &  t'   \\
t & r'
\end{array}
\right]\; ,
\label{S}
\end{equation}
where $r,r',t,$ and $t'$ are $2 \times 2$ reflection and transmission matrices, respectively. Thus,  
$\tau_1$ and $\tau_2$ in Eq. (\ref{concu_t1t2}) are the eigenvalues of the product $tt^\dagger$. In the presence of TRI, 
$S$ is unitary and symmetric. If TRI is broken (due to, e.g., the application of a magnetic flux), then $S$ is only unitary.
 
On the other hand, the chaotic scattering in the quantum dot gives a stochastic character to the entanglement production. Therefore, a statistical analysis of the entanglement is required. 
Previous works focused only on the mean value and variance 
of the concurrence. \cite{BKMY04,FMF06} As we have mentioned, here we obtain the complete distribution of the concurrence and entanglement of formation. In the framework of random matrix theory, the statistical properties of the $S$ matrix depend on the symmetry class $\beta$.\cite{carlo-review,pier-book} In the presence of TRI, the statistics of an ensemble of unitary and symmetric $S$ matrices is described by the so-called circular orthogonal ensemble ($\beta=1$). When TRI is absent, the statistical properties of $S$ are described by the Circular Unitary Ensemble ($\beta=2$). 
 Particular attention has been given in the past to the statistical properties of the transmission eigenvalues $\tau_n$ due to their relevance in quantum transport phenomena. In fact, the joint distribution $p_\beta(\{ \tau_n \})$ is known. Here we are interested in the joint distribution of only two eigenvalues, which is 
given by \cite{baranger-mello,jalabert}
\begin{equation}
\label{dist_taus}
p_{\beta}( \tau_1,\tau_2)=c_\beta| \tau_1 - \tau_2 |^{\beta}
(\tau_1 \tau_2)^{\beta/2-1},
\end{equation}
where $c_\beta$ is a normalization factor. 
Thus, the production of maximally or minimally entangled states is eventually determined by the statistical distributions 
of $\tau_1$ and $\tau_2$ through the relations (\ref{formation}) and (\ref{concu_t1t2}).
\begin{figure}
\includegraphics[width=\columnwidth]{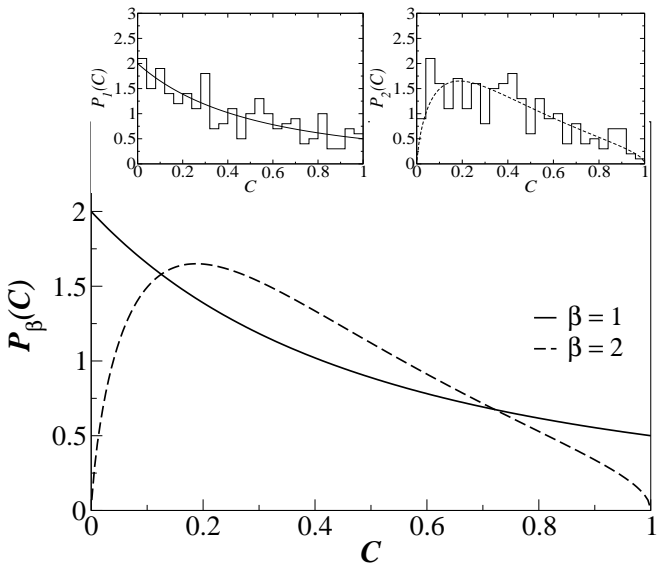}
\caption{Distribution of concurrence in the presence ($\beta=1$) and absence ($\beta=2$) of time reversal invariance. 
Insets: Numerical verification of the $P_\beta(\mathcal E)$. A good agreement between theory and numerics can be 
observed in both symmetry cases.
}
\label{fig-2}
\end{figure}

We can now calculate the distribution of the concurrence $P_\beta(\mathcal C)$ which is defined as  
\begin{equation}
\label{pofc}
P_\beta (\mathcal C)= \left< \delta\left[ \mathcal C-\frac{2\sqrt{\tau_1(1-\tau_1)\tau_2(1-\tau_2)}}
{\tau_1+\tau_2-2\tau_1\tau_2} \right] \right>_\beta ,
\end{equation}
where $\langle \cdots \rangle$ stands for the ensemble average performed with the probability
distribution of  Eq. (\ref{dist_taus}). 
The double integral over the variables $\tau_1$ and $\tau_2$  in Eq. (\ref{pofc}) can be 
performed exactly. After somewhat lengthy but direct algebra, we obtain a surprisingly simple result for the 
case of preserved TRI ($\beta=1$): 
\begin{equation}
\label{pofc_beta1}
P_1(\mathcal C)= \frac{2}{(1+\mathcal C)^2},
\end{equation}
whereas for broken TRI ($\beta=2$), we obtain the more cumbersome result:
\begin{eqnarray}
\label{pofc_beta2}
P_2(\mathcal C) &=& \frac{2\mathcal C}{(1-\mathcal C^2)^3}\left[3(2+3\mathcal C^2)\textrm{arctanh} 
\sqrt{1-\mathcal C^2} \right. \nonumber \\ && \left.-   (11+4 \mathcal C^2) \sqrt{1-\mathcal C^2} \right].
\end{eqnarray}
In Fig.~\ref{fig-2}, we plot the distributions $P_1(\mathcal C)$ (solid line) and $P_2(\mathcal C)$ (dashed line) 
as given by Eqs.~(\ref{pofc_beta1}) and (\ref{pofc_beta2}), respectively. The resulting curves look quite different. 
The probability of producing maximally entangled states ($\mathcal C=1$) is finite for $\beta = 1$, while for 
$\beta = 2$ such probability vanishes. Besides, separable states ($\mathcal C =0$) are produced with a maximum 
probability for $\beta = 1$, whereas for $\beta = 2$ the distribution goes to zero for such disentangled states.

As we noted previously, maximum entanglement ($\mathcal C =1$) requires $\tau_1=\tau_2$. 
Due to the ``repulsion'' factor $|\tau_1-\tau_2|^\beta$ in the joint distribution (\ref{dist_taus}), 
one might expect that $P_\beta(\mathcal C =1)=0$ for both $\beta =1$ and $2$. However, this is not the case for $\beta=1$ 
due to an exact cancellation of the repulsion factor  when performing the integrals over $\tau_1$ and $\tau_2$ in 
Eq. (\ref{pofc}). As a verification of our analytical results, we compare the distribution given by Eqs. (\ref{pofc_beta1}) and 
(\ref{pofc_beta2}) with numerical simulations of a chaotic quantum dot as performed in Ref.~\onlinecite{FMF06}. A good 
agreement between numerical and analytical results can be seen in the left and right insets of Fig. \ref{fig-2}.

From the distributions (\ref{pofc_beta1}) and (\ref{pofc_beta2}), we can calculate all the moments of $P_{\beta}(\mathcal C)$. In particular, for the mean value, we obtain
\begin{equation}
\label{averageC}
\langle \mathcal C \rangle =\left\{  
\begin{array}{ll}
\ln4-1 \approx 0.3863& \mbox{for $\beta=1$,}\\
4\pi({21 \pi}/{64}-1) \approx 0.3875   & \mbox{for $\beta=2$}, \\
\end{array}
\right. 
\end{equation}
while for the variance of the concurrence we have
\begin{equation}
\label{varianceC}
\mathrm{var}(\mathcal C) =\left\{  
\begin{array}{ll}
2[1-2(\ln 2)^2]\approx  0.0782  \ \ \ \  \mbox{for $\beta=1$,}& \\
\frac{\pi^2}{256}(88-21\pi)(21\pi-40)-22 \approx 0.0565   \\ \hspace{4.cm}  \mbox{for $\beta=2$ .} & \\
\end{array}
\right.
\end{equation}
These results are in agreement with those reported in  Refs.~\onlinecite{BKMY04} and \onlinecite{FMF06}. Both moments are practically independent of whether TRI is preserved or not.
From Fig.~\ref{fig-2}, however, it is clear that the first two moments are not sufficient for a complete characterization 
of the distributions. In fact, the fluctuations given by $\sqrt{\mathrm{var}(\mathcal C)}$ 
are of the order of  $\langle \mathcal C \rangle$.  
Seminal works\cite{BKMY04} based on the study of only $\langle \mathcal C\rangle_\beta$ and  
$\mathrm{var}_\beta(\mathcal C)$ arrived at preliminary conclusions suggesting that the breaking 
of TRI has no significant effects on the production of entanglement, which is in contrast to other quantum 
properties linked to transport such as the conductance. By studying the complete distribution,  
we find that this is not the case and we conclude that TRI has remarkable effects in the production of entanglement. 
\begin{figure}
\includegraphics[width=0.85\columnwidth]{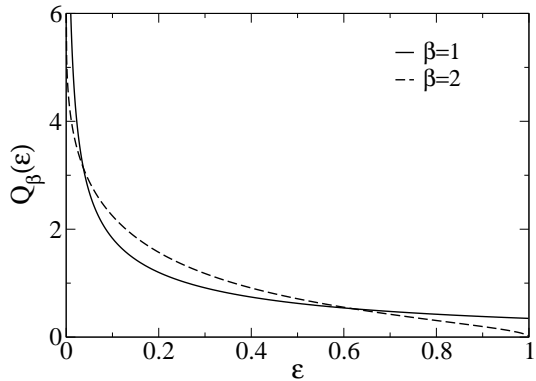}
\caption{Entanglement of formation distributions in the presence ($\beta=1$), and absence of time reversal invariance ($\beta=2$). 
}
\label{fig-3}
\end{figure}

Once we have calculated the distribution for the concurrence, we can easily obtain the distribution
for the entanglement of formation $Q_\beta(\mathcal E)$. We just need to make the change of variable 
$\mathcal C \to \mathcal E$ in Eqs. (\ref{pofc_beta1}) and (\ref{pofc_beta2}) by using Eqs. (\ref{formation}) and (\ref{h}). In this case, 
however, we can not give analytic expressions for $Q_\beta(\mathcal E)$ because of the transcendental functions in 
Eq. (\ref{h}). We proceed numerically, instead, by calculating the concurrence as a function of the entanglement of formation, $\mathcal C(\mathcal E)$. From the Jacobian of the transformation $\mathcal C \to \mathcal E$, we find that $Q_\beta(\mathcal E)$ is related to $P_\beta(\mathcal C)$ through  
\begin{equation}
 Q_\beta(\mathcal E)=\frac{1}{\mathcal C(\mathcal E)} \frac{\ln 2 \sqrt{1-C(\mathcal E) ^2}}{\mathrm{atanh} (\sqrt{1-\mathcal C(\mathcal E) ^2})}P_\beta(\mathcal C(\mathcal E)).
\end{equation}
The distribution $Q_\beta(\mathcal E)$ is depicted in Fig.~\ref{fig-3}. Of course, 
these curves give us the same information provided by $P_\beta(\mathcal C)$ in Fig.~\ref{fig-2},  and the 
interpretation of the TRI effects on the entanglement production is similar. However, 
$\mathcal E$ has the advantage of quantifying the cost of preparing an entangled state in terms of the 
amount Bell pairs needed. As regards the mean value and variance of $\mathcal E$, we obtain
\begin{equation}
\label{averageE}
\langle \mathcal E \rangle =\left\{  
\begin{array}{ll}
0.285 & \mbox{for $\beta=1$,}\\
 0.273   & \mbox{for $\beta=2$}, \\
\end{array}
\right. 
\end{equation}
and 
\begin{equation}
\label{varianceE}
\mathrm{var}(\mathcal E) =\left\{  
\begin{array}{ll}
0.078  & \mbox{for $\beta=1$,} \\
0.056   & \mbox{for $\beta=2$ .}  
\end{array}
\right.
\end{equation}
These two moments do not reveal any significant TRI effect, 
similar to what was obtained for the concurrence. 

Summarizing, we present a statistical analysis of the production of orbital entanglement in quantum-chaotic dots by calculating the full distributions of the corresponding  entanglement measures. Particularly, we found simple analytical expressions for the concurrence distributions in the presence and absence of TRI.  
The complete distributions of these measures allowed us to identify significant TRI  effects on the entanglement production, which are not reflected at the level of the first and second moments.

We thank S. Montangero for helpful discussions. This work was supported 
by the ``Ram\'on y Cajal" program of the Spanish Ministry of Education and Science.

%%%%%%%%%%%%%%%%%%%%%%%%%%%%%%%%%%%%%%%
%      BIBLIOGRAPHY
%%%%%%%%%%%%%%%%%%%%%%%%%%%%%%%%%%%%%%%

\end{document}